\documentclass[
 reprint,
superscriptaddress,
 amsmath,amssymb,
 aps,
 onecolumn
]{revtex4-2}

\usepackage{graphicx}
\usepackage{dcolumn}
\usepackage{bm}
\usepackage{makecell}
\usepackage{cases,mathtools}
\usepackage{float}
\usepackage{multirow}

\usepackage{array}    

\usepackage{xcolor}
\usepackage{hyperref}

\begin{document}

\preprint{APS/123-QED}

\title{Nonlinear tails of massive scalar fields around a black hole}

\author{Caiying Shao}\email[E-mail: ]{shaocaiying@ucas.ac.cn}\affiliation{School of Physical Sciences, University of Chinese Academy of Sciences,Beijing 100049, China}
\author{Zhen-Tao He}\email[E-mail: ]{hezhentao22@mails.ucas.ac.cn}\affiliation{School of Physical Sciences, University of Chinese Academy of Sciences,Beijing 100049, China}
\author{Jiageng Jiao}\email[E-mail: ]{jiaojiageng@ucas.ac.cn}\affiliation{International Centre for Theoretical Physics Asia-Pacific (ICTP-AP), University of Chinese Academy of Sciences (UCAS), Beijing, China.}\affiliation{Taiji Laboratory for Gravitational Wave Universe (Beijing/Hangzhou), University of Chinese Academy of Sciences (UCAS), Beijing, China.}
\author{Jingqi Lai}\affiliation{School of Physical Sciences, University of Chinese Academy of Sciences,Beijing 100049, China}
\author{Jun-Xi Shi}\affiliation{International Centre for Theoretical Physics Asia-Pacific (ICTP-AP), University of Chinese Academy of Sciences (UCAS), Beijing, China.}\affiliation{Taiji Laboratory for Gravitational Wave Universe (Beijing/Hangzhou), University of Chinese Academy of Sciences (UCAS), Beijing, China.}
\author{Yu Tian}\email[E-mail: ]{ytian@ucas.ac.cn}\affiliation{School of Physical Sciences, University of Chinese Academy of Sciences,Beijing 100049, China}
\author{Dandan Yuan}\affiliation{International Centre for Theoretical Physics Asia-Pacific (ICTP-AP), University of Chinese Academy of Sciences (UCAS), Beijing, China.}\affiliation{Taiji Laboratory for Gravitational Wave Universe (Beijing/Hangzhou), University of Chinese Academy of Sciences (UCAS), Beijing, China.}
\author{Hongbao Zhang}\email[E-mail: ]{hongbaozhang@bnu.edu.cn}\affiliation{School of Physics and Astronomy, Beijing Normal University, Beijing 100875, China}\affiliation{Key Laboratory of Multiscale Spin Physics, Ministry of Education, Beijing Normal University, Beijing 100875, China}
\date{\today}

\begin{abstract}
Nonlinear effects play a fundamental role in the late-time ringdown of black holes, with direct implications for gravitational-wave observations. 
For massive fields, these dynamics become richer, yet their nonlinear signatures remain poorly understood. 
Here, we systematically study nonlinear tails of massive scalar perturbations, from a toy model with ingoing and outgoing sources to a self-interacting scalar model, revealing nonlinear tails and contrasting the results with their linear counterparts. 
We find that the nonlinear tails of massive scalar fields, opposite to massless ones, decay as the same rate as linear tails at intermediate late times, independent of source parameters or initial conditions. 
Nevertheless, quadratic quasinormal modes could serve as a probe to the nonlinear effects of massive fields.
\end{abstract}

\maketitle

\section{\label{section1}Introduction}
In 2015, the LIGO–Virgo collaboration achieved the first successful detection of gravitational waves (GWs) from a compact binary merger~\cite{LIGOScientific:2016vlm}, marking a milestone that inaugurated a new era for testing general relativity and exploring the universe~\cite{prl-GW150914,prl-tests-gr,prl-constraints-dipole,Berti:2015itd,prd-theoretical-physics,Berti:2018cxi,Berti:2018vdi}. 
A key focus of current research is the analysis of black hole ringdowns, which encode both linear and nonlinear effects and leave distinctive imprints on the emitted GWs, providing unique insights into strong-gravity dynamics~\cite{Silva:2022srr,Shi:2019hqa,Shao:2023yjx}. 
With ongoing advances in technology, future space-based detectors, including LISA, Taiji, and Tianqin, are expected to detect ringdown signals from intermediate and supermassive black hole mergers with sufficiently high signal-to-noise ratios~\cite{amaroseoane2017laserinterferometerspaceantenna,Hu:2017mde,TianQin:2015yph,Gong:2021gvw}, opening a promising window for probing nonlinear aspects of black hole dynamics in previously inaccessible regimes.

Linear perturbation theory has long provided the foundation for understanding black hole ringdowns. 
It predicts that, after a merger, a black hole emits characteristic exponentially damped oscillations, known as quasinormal modes (QNMs), followed by a slower inverse power-law decay~\cite{Berti:2009kk,Price:1971fb,Price:1972pw,prd-qnm-lateti-Linear-01,Rosato:2025rtr,Hod:1999ci}. 
QNMs encode the characteristic ``sound" of a black hole, while the late-time tail carries distinctive signatures that differentiate black holes from the exotic compact objects and shed light on the structure of dynamical spacetimes~\cite{Cardoso:2016rao}.
Price first identified an inverse power-law tail $t^{-2l-3}$ for massless scalar perturbations in the Schwarzschild black hole spacetime~\cite{Price:1971fb}, a feature later shown to be generic for massless fields in spherically symmetric, asymptotically flat spacetimes~\cite{Ching:1995tj,Cardoso:2003jf}. 
However, massive fields decay slowly in an oscillatory  power-law manner~\cite{Hod:1998ra,Koyama:2001qw,Konoplya:2013rxa,Burko:2004jn,Chen:2005vq}, e.g., 
a scalar or gravitational field with mass $\mu$ decay as ${t^{ - l - 3/2}}\sin (\mu t)$ in the intermediate-time, highlighting the qualitatively different behavior from massless perturbations~\cite{Hod:1998ra}.
Not only are massive fields theoretically intriguing, but they are also of great significance in astronomical observations.
For instance, massive gravitons may constitute a previously overlooked source of long-wavelength gravitational waves, potentially contributing to the tails of ringdown signals, which could be probed by Pulsar Timing Arrays~\cite{Konoplya:2023fmh}.
In addition, massive bosonic fields can extract rotational energy from black holes via superradiance, forming dense boson clouds that oscillate and decay slowly~\cite{Konoplya:2024wds}.
The prolonged persistence of the tail of a massive field, relative to massless ones, increases the duration over which it can be observed, improving the chances of detecting its signal above the noise.

Recently, attention has shifted toward the nonlinear effects of massless perturbations around black holes~\cite{Okuzumi:2008ej,Mitman:2022qdl,Cheung:2022rbm,Kehagias:2025tqi,Ianniccari:2025avm,Alvares:2025pbi,DeAmicis:2024eoy,Ma:2024hzq}, including quadratic QNMs, whose amplitudes can be comparable to or even exceed those of linear QNMs, and nonlinear power-law tails, which decay more slowly than the standard Price tails~\cite{Kehagias:2025xzm,Cardoso:2024jme,Ling:2025wfv,He:2025ydh}.
While nonlinear tails agree with Price’s results for an inward-traveling source, for an outward-traveling source, their decay rates can, in certain cases, depend not only on the angular number $l$ but also on the source decay exponent $\beta$~\cite{Cardoso:2024jme,Ling:2025wfv}. 
Moreover, nonlinear 3+1 numerical relativity simulations indicate that these nonlinear features can dominate the late-time ringdown, underscoring their importance for accurate waveform modeling and potential detection in gravitational-wave observations~\cite{DeAmicis:2024eoy,Ma:2024hzq}.

While these nonlinear effects have been explored for massless perturbations~\cite{Berti:2025hly}, studies on massive nonlinear tails remain largely unexplored.
This naturally raises the question of how nonlinear tails behave for massive fields: whether they exhibit qualitative differences from the linear case, and to what extent the properties of the perturbation source influence them.
To address these questions, we intend to initiate a systematic study by analyzing the nonlinear tails of massive scalar fields around a black hole.
We first consider a toy model, where the nonlinear tails of massive fields are driven by outgoing or ingoing wavepackets. 
We then extend our study to a cubic self-interacting scalar field model, which naturally provides a quadratic coupling in the sources. 
We analyze the differences between intermediate-time tails and quadratic QNMs relative to their linear counterparts, finding that the nonlinear corrections to the tails of massive fields are less significant than in the massless case. 
Consequently, linear theory provides a sufficiently accurate approximation for the intermediate-time tails of the ringdown signal.
Furthermore, the quadratic QNMs may produce potentially observable signatures.
To the best of our knowledge, this study is the first to numerically investigate nonlinear tails and quadratic QNMs in massive scalar fields, using a highly efficient and robust double-null evolution scheme.

The remainder of this paper is organized as follows. 
In Section~\ref{section2}, we present the theoretical framework for investigating nonlinear tails of massive scalar fields, with two types of nonlinear sources: one is a moving toy model composed of outgoing and ingoing wavepackets, and the other originates from a cubic self-interacting scalar theory.
In Section~\ref{section3}, we present the numerical methods and key results, characterizing the dynamical evolution of a neutral massive scalar field, analyzing its intermediate-time tails across various source parameters, initial conditions, and related settings, and highlighting the quadratic QNMs as a potentially observable feature.
Finally, the concluding remarks are presented in Section~\ref{section4}.

\section{\label{section2}Theoretical framework of nonlinear tail dynamics}
In this section, we introduce the theoretical framework for studying nonlinear tails of massive scalar perturbations around a Schwarzschild black hole with mass $M$. 
We will set $c=G=2M=1$ in the following text.
We consider the wave equation of a massive scalar field $\Psi $ with mass $\mu$:
\begin{equation} \label{wave}
({\nabla ^a}{\nabla _a} - {\mu ^2})\Psi   = S,
\end{equation}
including a source term $S$ of the form 
\begin{equation}\label{betaS}
S = \sum\limits_{(lm)}  -  \frac{1}{{{r^{\beta  + 1}}}}F\left[ {{U_s}\left( {t - {t_i}} \right) - \left( {x - {x_i}} \right)} \right]{Y_{lm}},\quad \beta  \ge 0,\quad f(r) = 1 - \frac{1}{r},
\end{equation}
where a wavepacket $F$, with a velocity ${U_s}(0 < |{U_s}| \le 1)$ in the tortoise coordinate $x=\int 1/f dr $, encodes the dynamics of compact objects near black holes.
For ${U_s}>0$, it describes outward-propagating bursts, such as those characteristic of gamma-ray bursts~\cite{Birnholtz:2013bea}, while more complex trajectories can give rise to long-lived transients~\cite{Albanesi:2023bgi,DeAmicis:2024eoy}. 
Conversely, for ${U_s}<0$, the wavepacket represents inward-moving perturbations, such as those occurring in extreme-mass-ratio systems~\cite{amaro2018relativistic,babak2017science,Zi:2024lmt,barack2018self}.
Alternatively, the source can arise from a cubic self-interacting scalar theory,
\begin{equation}\label{lambdaS}
S = \frac{\lambda }{2}{\Psi^2},
\end{equation}
providing an ideal setting to study sourced tails.
In the Eq. \eqref{betaS}, we employ real spherical harmonics ${Y_{lm}}$, which are defined in terms of the standard complex spherical harmonics $Y_l^m$ as
\begin{equation}\label{N15}
{Y_{lm}} = 
\begin{dcases}
\frac{i}{\sqrt 2 }
\left[ {Y_l^m - {{( - 1)}^m}Y_l^{ - m}} \right]
& m<0
\\
Y^0_l & m=0
\\
\frac{1}{\sqrt 2}
\left[ {Y_l^{ - m} + {{( - 1)}^m}Y_l^m} \right] & m > 0
\end{dcases}.
\end{equation}

Expanding the field as
\begin{equation}\label{N15}
\Psi (t,r,\theta ,\varphi ) = \frac{1}{r}\mathop \sum \limits_{l = 0}^\infty  \mathop \sum \limits_{m =  - l}^l {\phi _{lm}}(t,r){Y_{lm}}(\theta ,\varphi ),
\end{equation}
substituting the Schwarzschild metric into the Eq. \eqref{wave}, and projecting onto the real spherical harmonic basis, one can obtain a set of (1+1)-dimensional partial differential equations, each governing the evolution of a specific $(l,m)$ mode:
\begin{equation}\label{N15qwe}
\left[ {\partial _t^2 - \partial _x^2 + V_l\left( x \right)} \right]{\phi _{lm}} =
S_{lm}=
\begin{dcases}
&\frac{{f(r)}}{{{r^\beta }}}F\left[ {U_s}\left( t - t_i \right) - \left( {x - {x_i}} \right)
\right]
\\
&
\frac{{\lambda 
(1 - r)
}}{{2{r^2}}}\sum\limits_{({l_1}{m_1})} {\sum\limits_{({l_2}{m_2})} {C_{{l_1}{m_1},{l_2}{m_2}}^{lm}{\phi _{{l_1}{m_1}}}{\phi _{{l_2}{m_2}}}} },
\end{dcases}
\end{equation}
with an effective potential
\begin{equation}\label{N15}
V_l\left( x \right) = \frac{{f(r)\left[ {l(l + 1) + {\mu ^2}{r^2} + rf'(r)} \right]}}{{{r^2}}}.
\end{equation}

For the source term Eq. \eqref{lambdaS}, the nonlinear coupling coefficients ${C_{{l_1}{m_1},{l_2}{m_2}}^{lm}}$ arise from the angular integrals of real spherical harmonics
\begin{equation}\label{N15ert}
C_{{l_1}{m_1},{l_2}{m_2}}^{lm} = \int {d\Omega } {Y_{{l_1}{m_1}}}(\theta ,\varphi ){Y_{{l_2}{m_2}}}(\theta ,\varphi ){Y_{lm}}(\theta ,\varphi ).
\end{equation}
Taking into account the nonlinear term $\lambda {\Psi ^2}/2$, the solution can be represented as a power series in $\lambda$ in the small-$\lambda$ regime
\begin{equation}\label{N15x}
{\phi _{lm}}\left( {t,x} \right) = \mathop \sum \limits_{n = 0}^\infty  {\lambda ^n}\phi _{lm}^{(n)}\left( {t,x} \right),
\end{equation}
where $n$ denotes the perturbative order of ${\phi _{lm}}$.
In particular, for $n=0$, one naturally recovers the solution of the linear Regge-Wheeler equation.
By substituting this perturbative expansion~\eqref{N15x} into Eq. \eqref{N15qwe}, the corresponding component $\phi _{lm}^{(n)}$ at each perturbative order $n$ is determined by the following equation:
\begin{equation}\label{N15}
{\left( {\partial _t^2 - \partial _x^2 + {V_l}} \right)}
{\phi _{lm}^{(n)}} = S_{lm}^{(n)},
\end{equation}
with 
\begin{equation}\label{N15}
S_{lm}^{(n)} = \frac{{(1 - r)}}{{2{r^2}}}
\sum_{({l_1}{m_1})}
\sum_{({l_2}{m_2})}
\sum_{({n_1}{n_2})}
{C_{{l_1}{m_1},{l_2}{m_2}}^{lm}{\phi^{(n_1)} _{{l_1}{m_1}}}
{\phi^{(n_2)} _{{l_2}{m_2}}}} ,
\end{equation}
where the summation of $({n_1}, {n_2})$ is subject to $n_1+n_2=n-1$.
In this paper, we consider only the second-order perturbation, corresponding to $n=1$.

\section{\label{section3}Numerical methods and relevant results}
In this section, we employ the finite difference method~\cite{prd-qnm-lateti-Linear-01} to numerically evolve the scalar field.
We begin by performing a coordinate transformation to double-null coordinates, defined as {$u = t - {x}$ and $v = t + {x}$}, which allows the Eq. \eqref{N15qwe} to be recast in the form
\begin{equation}\label{N15}
\frac{{{\partial ^2}\phi }}{{\partial u\partial v}} + \frac{{V[r(u,v)]\phi }}{4} = S\left( {u,v} \right).
\end{equation}
Applying the finite difference scheme, the field value at the grid point N is computed from the neighboring points W, E, and S as
\begin{equation}\label{N16as}
{\phi _\text{N}} = {\phi _\text{W}} + {\phi _\text{E}} - {\phi _\text{S}} - \Delta u\Delta vV[r(u,v)]\frac{{{\phi_\text{W}} + {\phi _\text{E}}}}{8} + \Delta u\Delta vS\left( {u,v} \right).
\end{equation}
Here, the indices N, W, E, and S correspond to the grid points N $ \equiv (u + \Delta u,v + \Delta v)$, W$\equiv (u,v + \Delta v)$, E $\equiv (u + \Delta u,v)$, and S $\equiv (u,v)$ where $\Delta u$ and $\Delta v$ denote the step sizes in $u$ and $v$, respectively.  

For source-free or linear perturbations, the time-domain profile is generated by specifying the initial conditions
\begin{equation}\label{N17}
\phi (u,0) = 0,\quad \phi (0,v) = \exp\left[ 
- \frac{{{{\left( {v - {v_c}} \right)}^2}}}{{2{\sigma ^2}}}
\right],
\end{equation}
where ${v_c}$ and $\sigma $ denote the center and width of a Gaussian pulse specified on the initial ingoing null segment $u=0$.
Physically, this corresponds to characteristic initial data describing a localized perturbation injected from the null boundary.
It is therefore localized on the initial null slice, rather than on a constant-$t$ spatial slice, which is a standard choice in double-null evolutions.
In contrast, for source-driven or nonlinear perturbations, we adopt vanishing initial conditions,
\begin{equation}\label{N17nl}
\phi(u,0) = 0, \quad \phi(0,v) = 0,
\end{equation}
so that the evolution is entirely driven by the source term.

For the numerical study of the toy model, one may further define 
\begin{equation}\label{N17}
F(x) = \frac{1}{{\sqrt {2\pi } {\sigma _F}}}\exp \left( { -\frac{{ {x^2}}}{{2\sigma _F^2}}} \right),
\end{equation}
As ${\sigma _F} \to 0$, $F(x)$ approaches a Dirac delta function, allowing us to approximate the source as a collection of pointlike particles.
For the self-interacting scalar field, we analyze the dynamics at second order, initializing only the first-order $(l,m)=(1,1)$ mode while leaving all other modes unexcited. 
The nonlinear coupling then generates second-order contributions in other modes, which are explicitly given by
\begin{equation}\label{N17}
S_{00}^{(1)} = \frac{{ (1 - r)}}{{2{r^2}}}\left[ {\frac{{{{\left( {\phi _{11}^{(0)}} \right)}^2}}}{{2\sqrt \pi  }}} \right],\quad S_{20}^{(1)} = \frac{{ (1 - r)}}{{2{r^2}}}\left[ { - \frac{{{{\left( {\phi _{11}^{(0)}} \right)}^2}}}{{2\sqrt {5\pi } }}} \right],\quad S_{22}^{(1)} = \frac{{ (1 - r)}}{{2{r^2}}}\left[ {\frac{1}{2}\sqrt {\frac{3}{{5\pi }}} {{\left( {\phi _{11}^{(0)}} \right)}^2}} \right].
\end{equation}
Finally, the full temporal evolution $\phi (t,x)$ is obtained from the numerically propagated data at successive time steps, providing the complete profile of the field.
 
\begin{figure*}[htbp]
\centering
\includegraphics[scale=0.45]{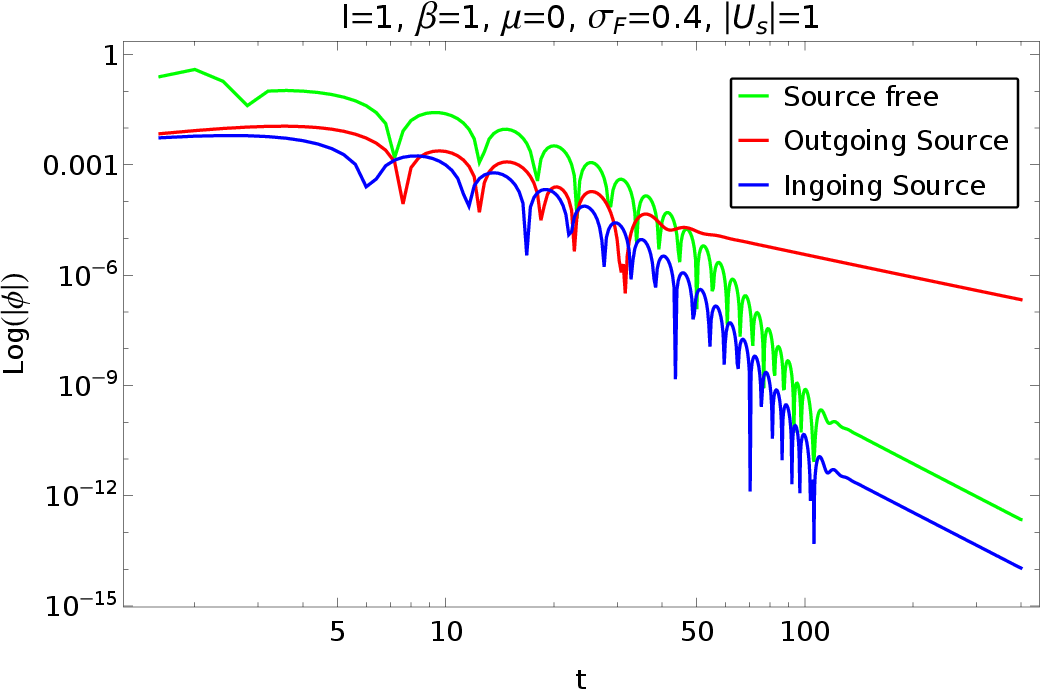}
\caption{Time evolution of massless scalar perturbations, comparing outgoing and ingoing cases with a moving source to the source-free case.
}\label{Fig1}
\end{figure*}

\begin{figure*}[htbp]
\centering
\includegraphics[scale=0.45]{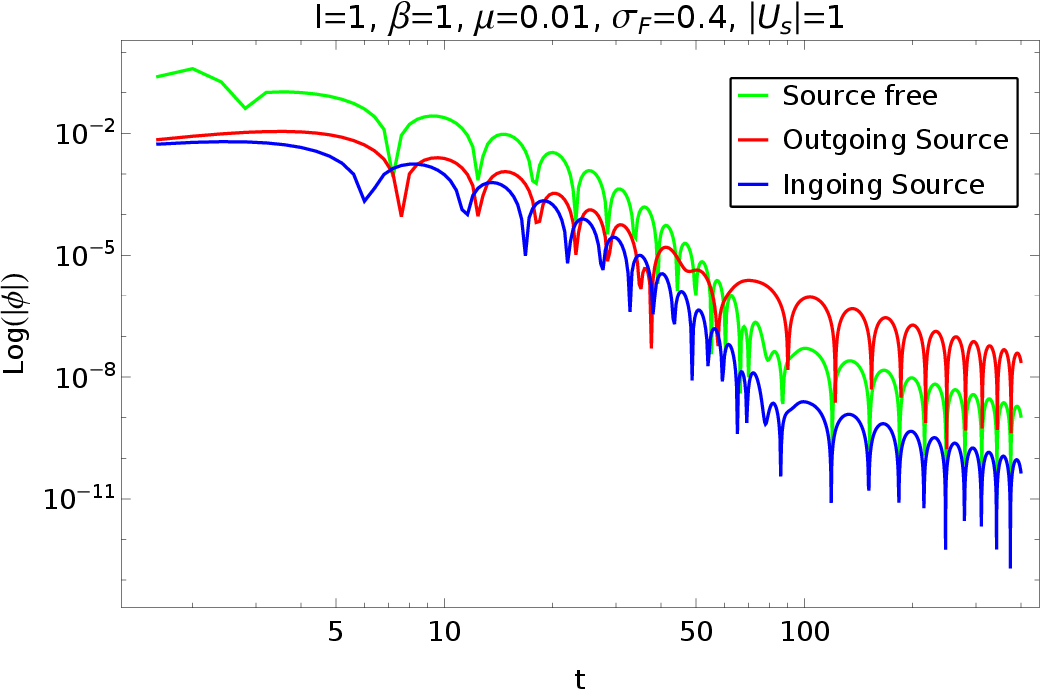}
\includegraphics[scale=0.45]{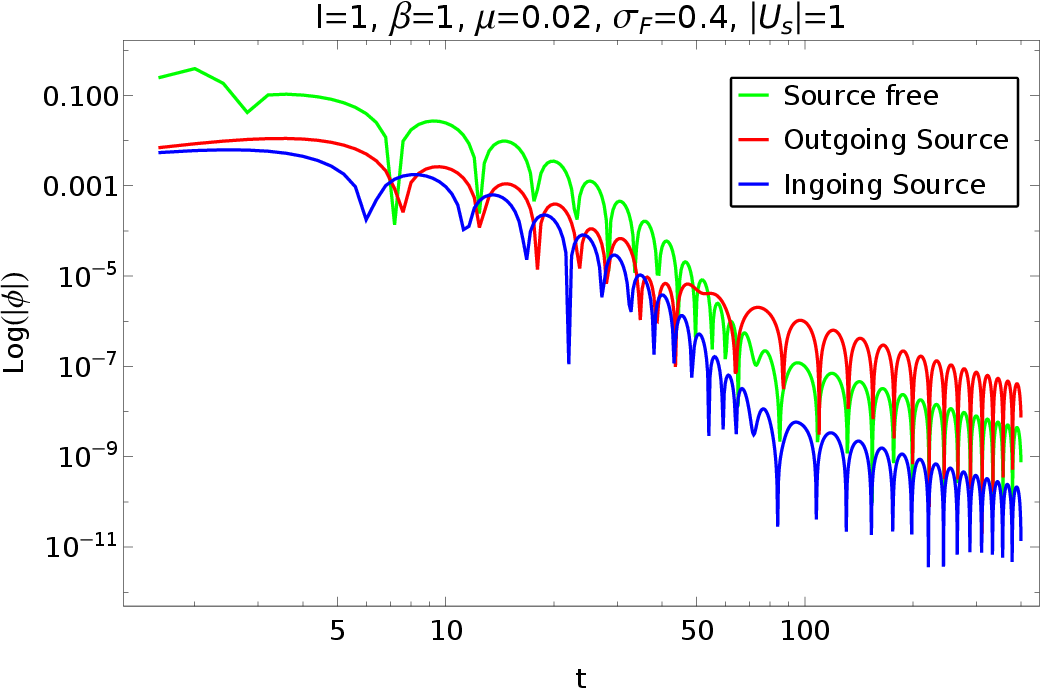}
\includegraphics[scale=0.45]{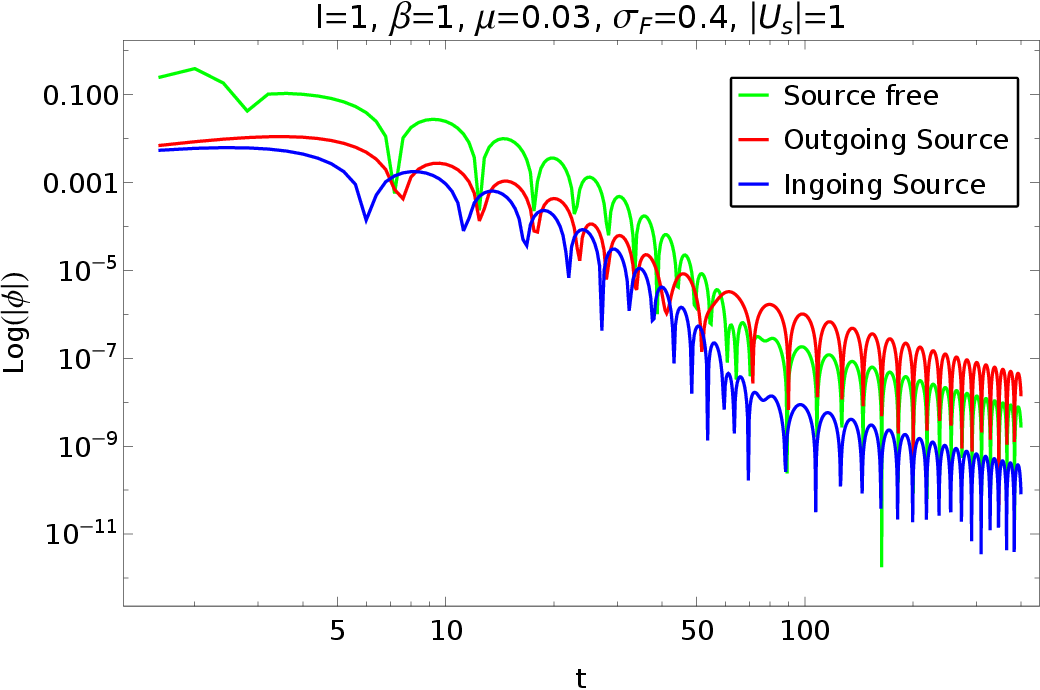}
\includegraphics[scale=0.45]{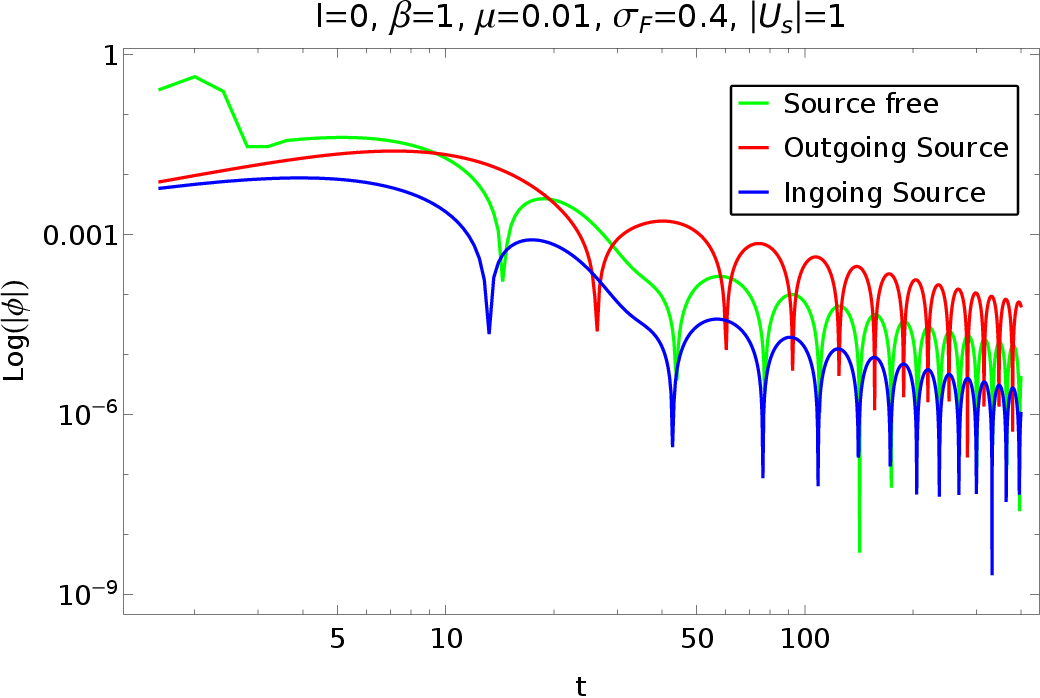}
\includegraphics[scale=0.45]{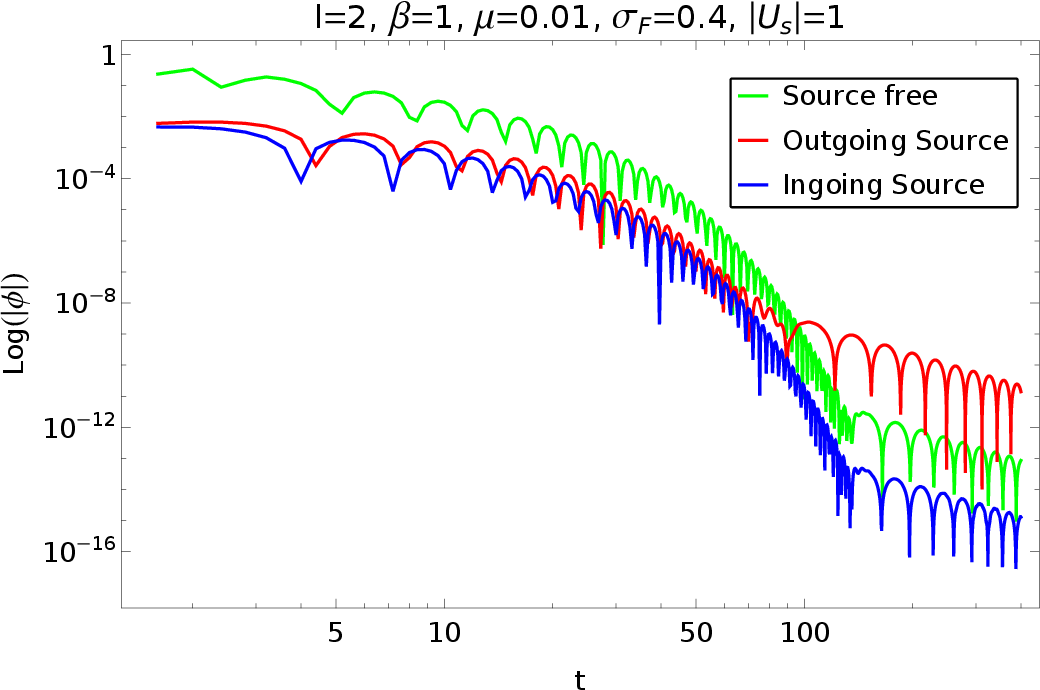}
\includegraphics[scale=0.45]{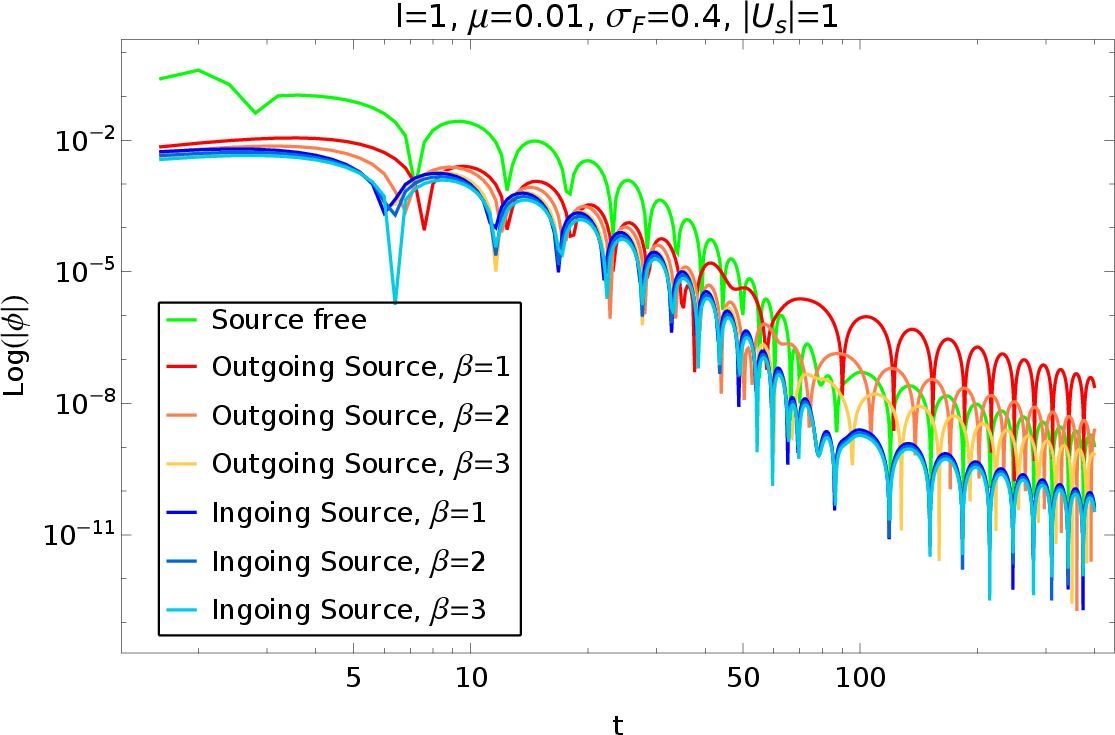}
\includegraphics[scale=0.45]{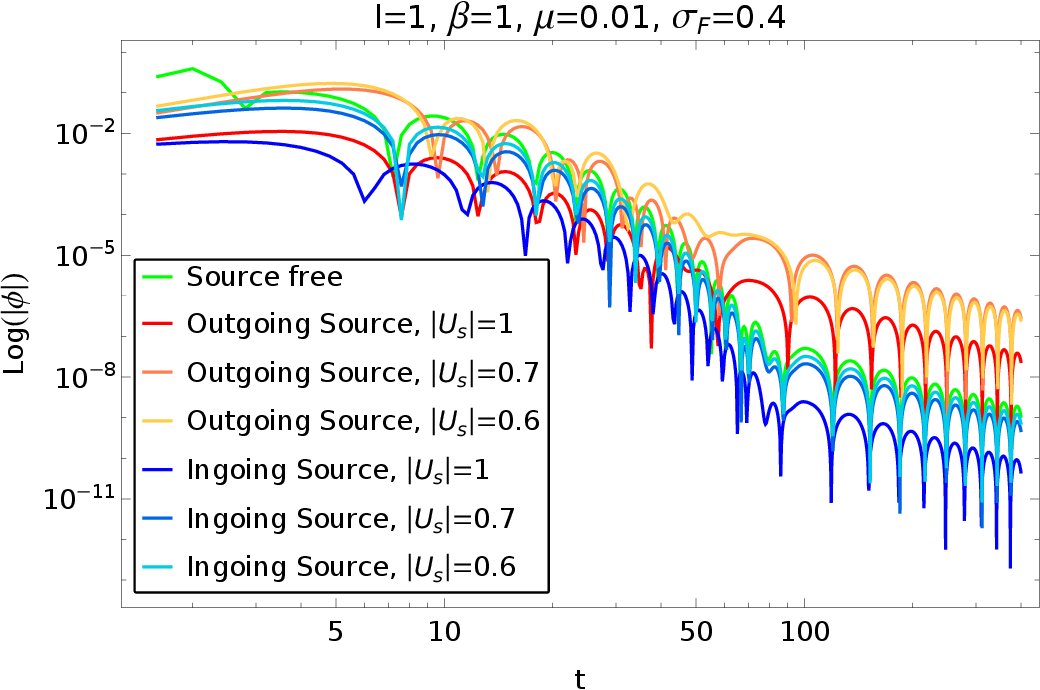}
\includegraphics[scale=0.45]{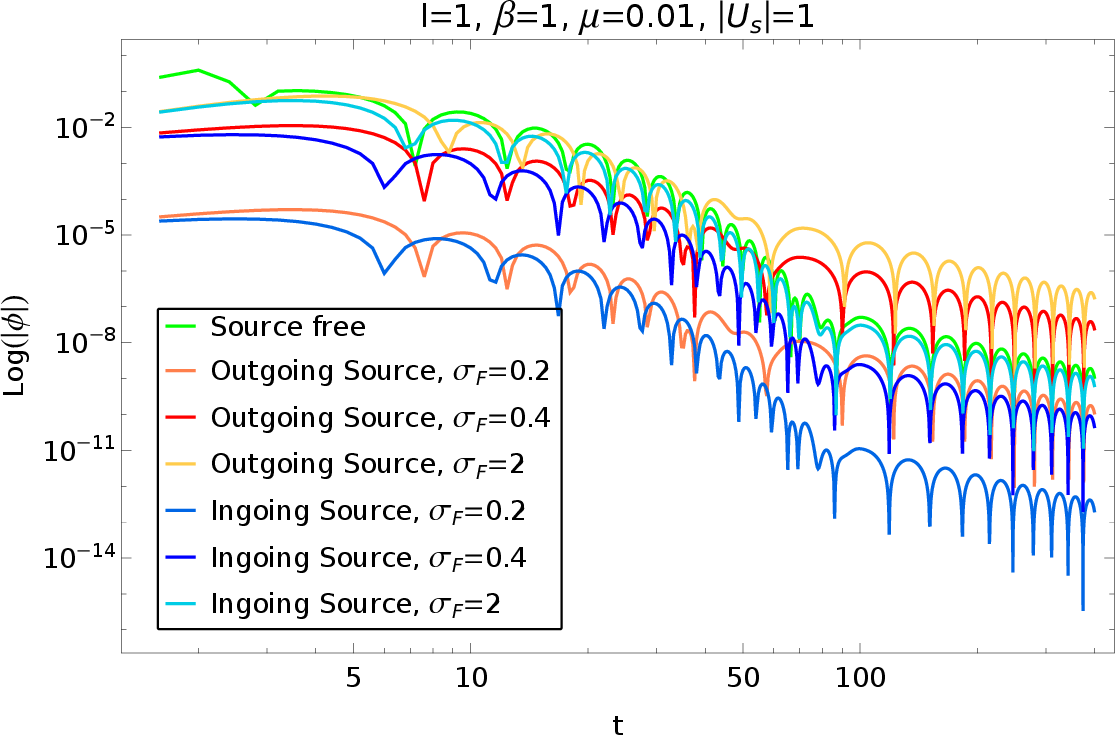}
\caption{Quasinormal oscillations and tails of massive scalar perturbations for both source-free and source-driven cases, including outgoing and ingoing wavepackets.
}\label{Fig2}
\end{figure*}

For the toy model analysis of nonlinear tails, we first consider the case of a massless scalar field, as illustrated in Fig.~\ref{Fig1}. 
For an inward-propagating source with ${U_s}=1$, the field exhibits the expected Price tails. 
In contrast, for an outward-propagating source, the nonlinear tails decay more slowly for $\beta=1$ compared to the tails predicted by linear theory, consistent with previous studies \cite{Cardoso:2024jme, Ling:2025wfv}.
However, as the mass of the scalar field increases, oscillations emerge in the tails.
As shown in Fig.\ref{Fig2}, we find that the source-related parameters, namely the power-law exponent $\beta $, the standard deviation ${{\sigma _F}}$, and the source velocity ${U_s}$, have little to no effect on the decay rate of the nonlinear tails, regardless of whether the source is inward- or outward-propagating.
This behavior contrasts sharply with the massless case, where the decay of nonlinear tails depends on the propagation direction of the source and, in certain cases, the tail decay rate is determined not only by the angular quantum number $l$ but also by $\beta$~\cite{Cardoso:2024jme,Ling:2025wfv}.
Across the intermediate-time tail of the ringdown signal, we observe that increasing the source parameter $\beta$ reduces the amplitude of the tails, whereas increasing ${{\sigma _F}}$ leads to an increase in amplitude.
These results suggest that, for a massive scalar field, nonlinear effects play a less significant role in tail decay compared to the massless case.
Consequently, linear theory provides a good approximation for the majority of the tail signal, with the intermediate-time decay primarily determined by the angular quantum number $l$ and the oscillation frequency mainly set by the scalar field mass $\mu $. 

\begin{figure*}[!]
\centering
\includegraphics[scale=0.45]{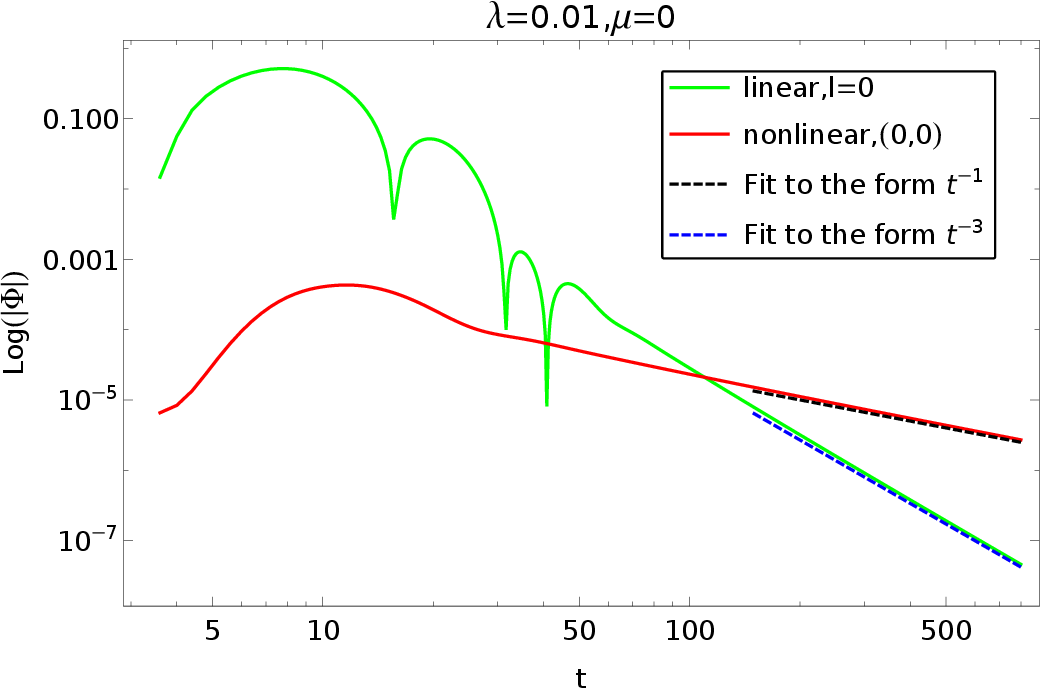}
\includegraphics[scale=0.45]{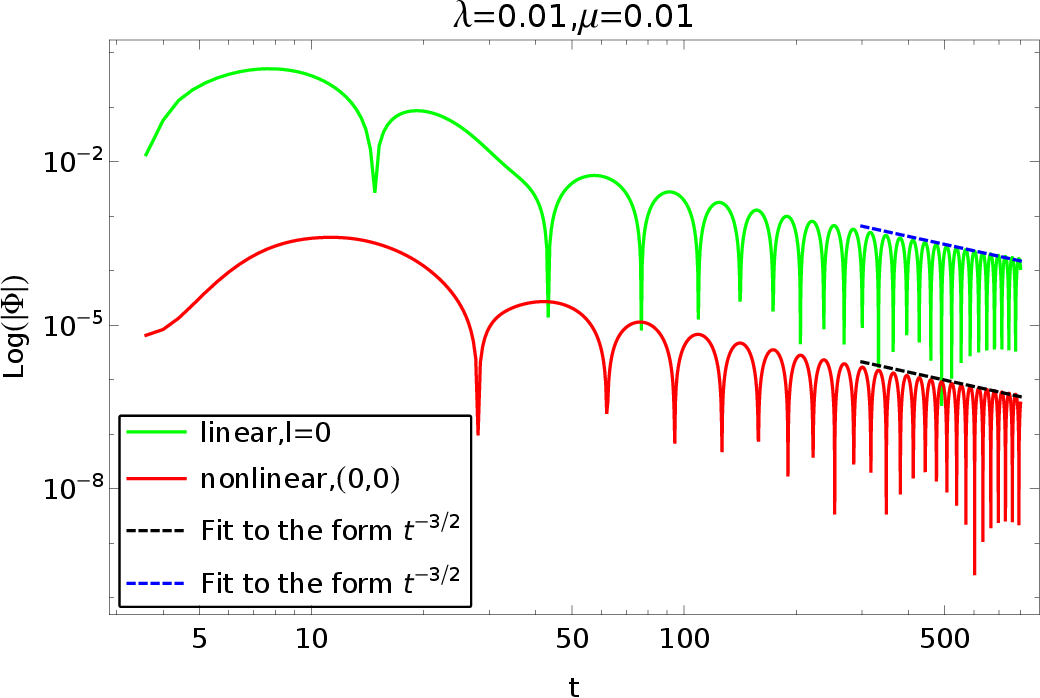}
\caption{Temporal evolution of massless and massive scalar perturbations for linear and nonlinear cases. 
The evolved field has been rescaled as ${\Phi _{lm}} = {\lambda ^n}\phi _{lm}^{(n)}$.
For the massless case, dashed lines show analytical predictions for both linear and nonlinear decay, in excellent agreement with the numerical data. 
For the massive case, dashed lines indicate linear analytical predictions for the intermediate-time tails only, yet they accurately capture both linear and nonlinear numerical evolution.
}\label{Fig3}
\end{figure*}

\begin{figure*}[!]
\centering
\includegraphics[scale=0.45]{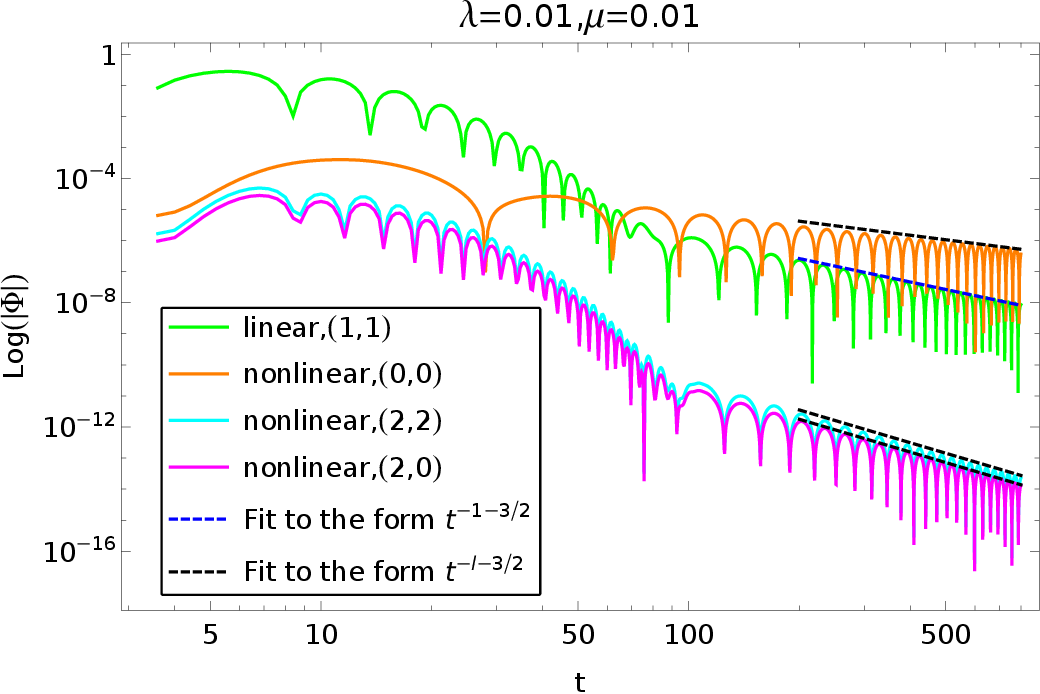}
\includegraphics[scale=0.45]{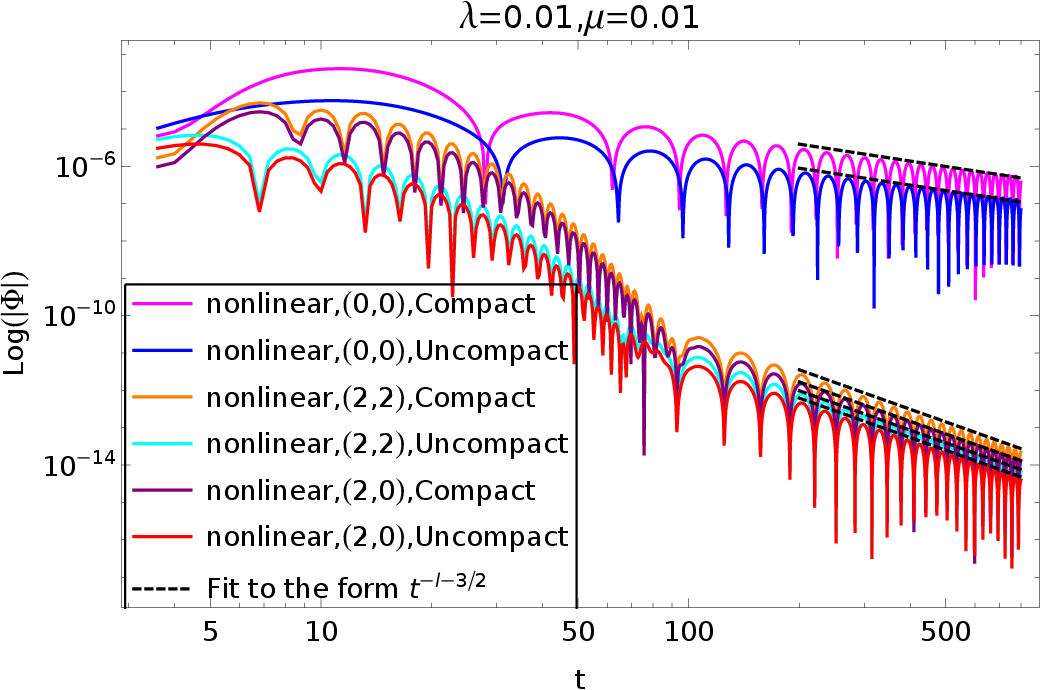}
\caption{Temporal evolution of massive scalar perturbations, calculated for the linear $(1,1)$ mode and three nonlinear modes. 
The evolved field has been rescaled as ${\Phi _{lm}} = {\lambda ^n}\phi _{lm}^{(n)}$.
The nonlinear modes $(0,0)$, $(2,2)$, and $(2,0)$ are generated by self-couplings of the $(1,1)$ mode. 
Linear analytical predictions for the intermediate-time tails (dashed lines) remain in excellent agreement with the numerical evolution, even when nonlinear couplings are present. 
Changing the initial data type (compact vs. uncompact) affects only the overall amplitude, while the numerical results remain in excellent agreement with the linear analytical predictions.
}\label{Fig4}
\end{figure*}

Although toy models offer useful insights into the fundamental behavior of nonlinear tails, these simplified source models often contain artificial elements. 
A more natural and ideal approach involves self-interacting scalar field systems, where mode couplings and nonlinear terms naturally arise, shaping the evolution of the field.
From this perspective, we focus on the $(1,1)$ linear mode, which undergoes self-coupling and gives rise to nonlinear excitations in the $(0,0)$, $(2,0)$, and $(2,2)$ modes.
As shown in Fig.~\ref{Fig3}, by comparing the nonlinear $(0,0)$ mode with the linear $l=0$ mode, we observe that for a massless scalar field, the linear tail follows the Price law~\cite{Price:1971fb}, whereas the nonlinear tail decays more slowly, following a power-law behavior of the form ${t^{ - l - \beta }}$, with $\beta  = 1$ and $l = 0$~\cite{Ling:2025wfv}. 
Upon introducing a finite scalar field mass, the decay rates of linear and nonlinear tails converge, exhibiting no significant distinction in their intermediate-time behavior.
For comparison, we consider the linear $(1,1)$ mode and the nonlinear $(0,0)$, $(2,0)$, and $(2,2)$ modes, as shown in the left panel of Fig.~\ref{Fig4}, and find that the intermediate-time tails of both linear and nonlinear modes decay consistently with the analytical prediction for linear intermediate-time tails, ${t^{ -l-3/2}}$~\cite{Hod:1998ra}.

We vary the type of initial perturbation to compare the responses for compact and non-compact configurations, and observe that these modifications primarily affect the amplitude of the nonlinear tails, while the decay rate remains unchanged (see the right panel of Fig.~\ref{Fig4}).

\begin{figure*}[!]
\centering
\includegraphics[scale=0.5]{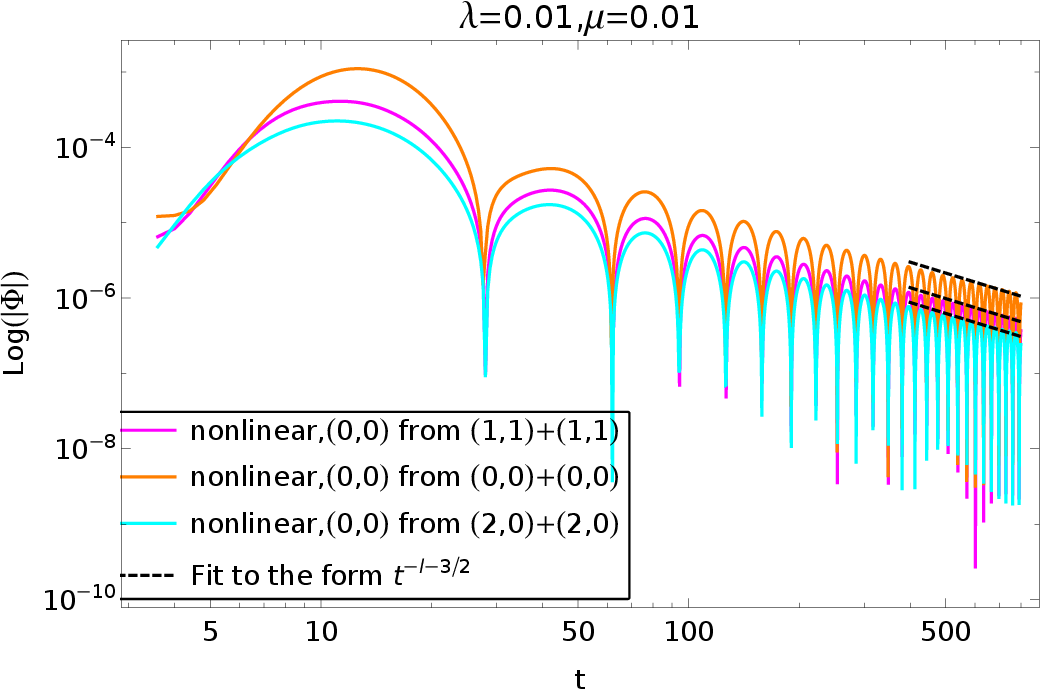}
\caption{Temporal evolution of massive scalar perturbations in nonlinear simulations. 
The evolved field has been rescaled as ${\Phi _{lm}} = {\lambda ^n}\phi _{lm}^{(n)}$.
The $(0,0)$ mode is generated via quadratic self-couplings of the linear $(1,1)$, $(0,0)$, and $(2,0)$ modes. 
Dashed lines denote linear analytical predictions for the intermediate-time tails, which are in excellent agreement with the corresponding nonlinear numerical evolution.
}\label{Fig5}
\end{figure*}

In addition, we also examine how different coupling channels affect the nonlinear tail. 
Assuming that the initial data consist of only the $(1,1)$ mode, only the $(0,0)$ mode, or only the $(2,0)$ mode, we find that self-coupling in each case can generate the $(0,0)$ mode, as described by
\begin{equation}\label{N17}
S_{00}^{(1)} = \frac{{ (1 - r)}}{{2{r^2}}}
\left[ {\frac{{{{\left( {\phi _{11}^{(0)}} \right)}^2}}}{{2\sqrt \pi  }}} \right],
\quad S_{00}^{(1)} = \frac{{ (1 - r)}}{{2{r^2}}}
\left[ {\frac{{{{\left( {\phi _{20}^{(0)}} \right)}^2}}}{{2\sqrt \pi  }}} \right],\quad S_{00}^{(1)} = \frac{{ (1 - r)}}{{2{r^2}}}\left[ {\frac{{{{\left( {\phi _{00}^{(0)}} \right)}^2}}}{{2\sqrt \pi  }}} \right].
\end{equation}
Fig. \ref{Fig5} illustrates the self-coupling of the three channels corresponding to the linear $(1,1)$, $(0,0)$, and $(2,0)$ modes. 
We observe that, regardless of the specific combinations of these channels, the nonlinear tails decay in accordance with the linear analytical solution.

Our results indicate that the nonlinear decay rates of intermediate-time tails remain robust, surprisingly consistent with their linear counterparts.
In linearized perturbation theory, the intermediate-time tail behaviors depend on the scalar field mass $\mu$, independent of the backscattering off the spacetime curvature from far regions \cite{Hod:1998ra}.
Given this picture, we conjecture that the backscattering induced by the nonlinearities could also be neglected at the intermediate late times, and might influence the asymptotic late-time tail behaviors.
Investigating the nonlinear asymptotic late-time tail behaviors numerically are more challenging, which, as well as a formal analysis via the Green's function method, warrants further researches. 

\begin{figure*}[!]
\centering
\includegraphics[scale=0.5]{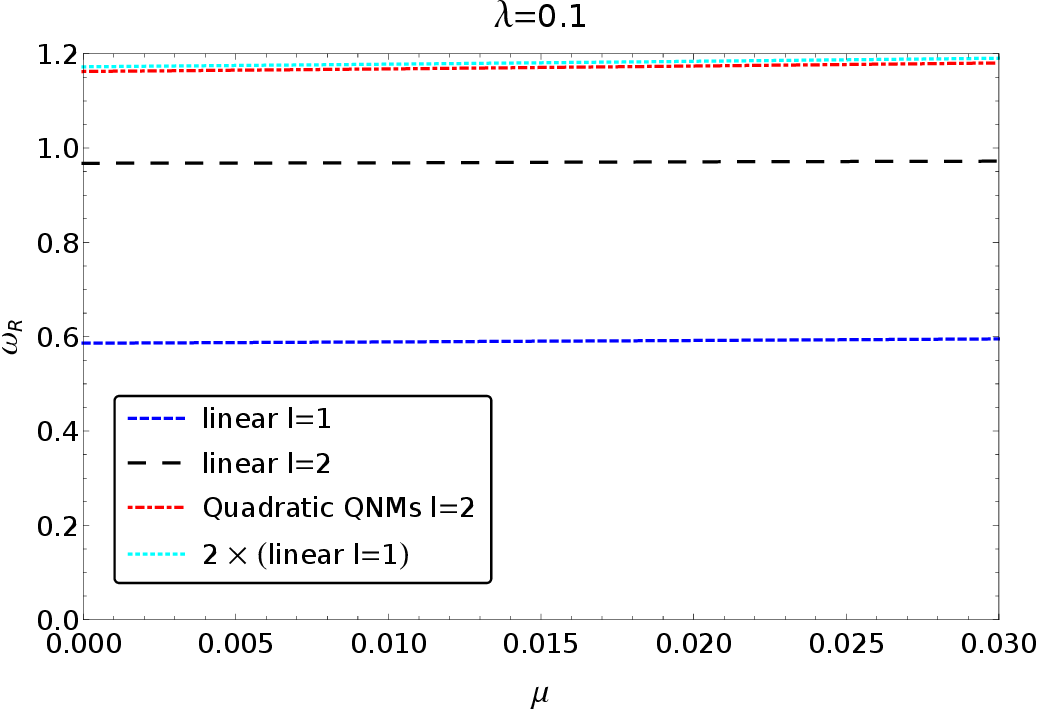}
\includegraphics[scale=0.5]{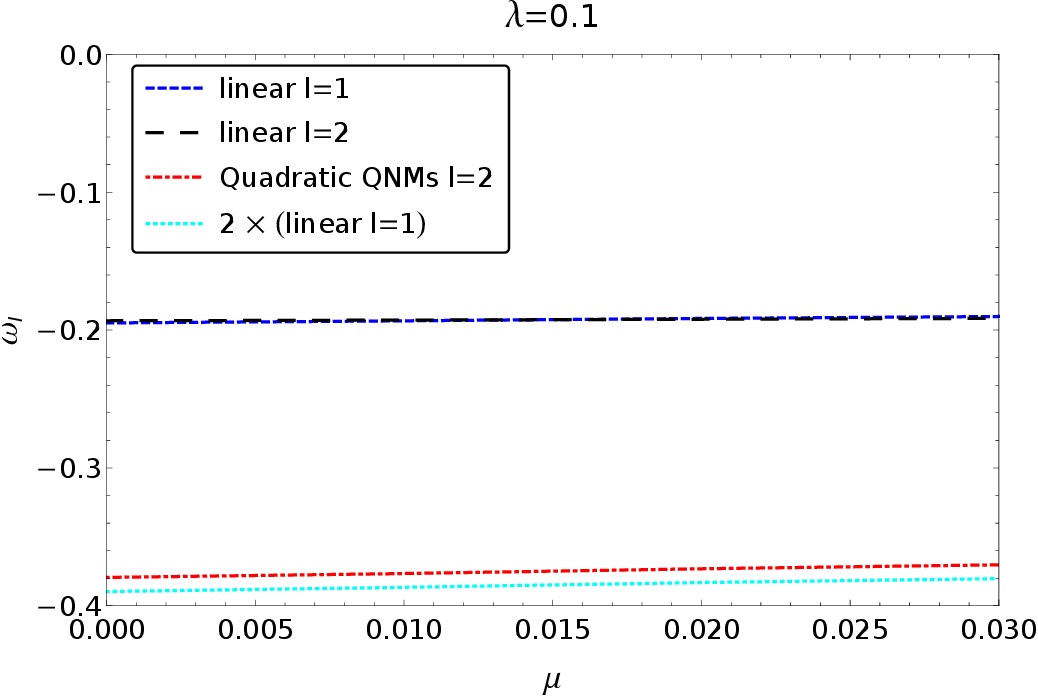}
\caption{Real and imaginary parts of the lowest-lying quasinormal modes $\omega=\omega_R+i\omega_I$ as functions of the scalar-field mass $\mu$.
The blue dashed line denotes the linear $l=1$ mode, the black dashed line represents the linear $l=2$ mode, the red dot-dashed line corresponds to the quadratic QNMs extracted from the second-order $l=2$ channel generated by the self-coupling of the linear $(1,1)$ mode, and the cyan dotted line indicates twice the frequency of the linear $l=1$ mode.
}\label{Fig6}
\end{figure*}

While the intermediate-time tail of a massive scalar perturbation remains largely insensitive to nonlinear effects, the quasinormal mode oscillations provide a direct probe of nonlinearity. 
To illustrate this, we analyzed the QNMs of self-interacting scalar fields using the matrix pencil method \cite{Berti:2007dg}, focusing on the lowest-lying modes as the scalar field mass varies. 
The results are shown in Fig.~\ref{Fig6}, where the real and imaginary parts of the extracted frequencies are plotted as functions of $\mu $.
The figure displays the linear $l=1$ and $l=2$ modes, together with the quadratic-QNM contribution extracted from the second-order $l=2$ channel generated by the self-coupling of the linear $(1,1)$ mode.
The second-order $l=2$ waveform in fact contains two oscillatory components: an inherited contribution at the linear $l=2$ QNM frequency, and a genuine quadratic contribution generated by the nonlinear source.
Theoretically, quadratic QNMs occurs exactly at twice the corresponding first-order frequency. 
The small numerical deviations, as seen in Fig.~\ref{Fig6} for the cyan dotted and red dash-dotted lines, primarily arise from numerical errors from the finite difference \eqref{N16as} and finite fitting windows.
This observation is consistent with the original second-order analysis of Schwarzschild ringdown, as well as with recent studies on quadratic QNMs and their application in nonlinear tests of gravity~\cite{Nakano:2007cj, Bucciotti:2024zyp, Yi:2024elj, Lagos:2024ekd}.
This also implies that higher-order perturbative effects might play an important role in understanding the full dynamics of black hole ringdown.
Although the nonlinear modes (2,2) and (2,0) are difficult to distinguish based on their frequencies alone, they can be separated by comparing their amplitudes. 

\section{\label{section4}Concluding remarks}
In this work, we investigate massive scalar perturbations in the Schwarzschild spacetime by considering two complementary setups: a toy model that tracks the dynamical propagation of ingoing and outgoing wavepackets, and a self-interacting scalar field system that incorporates nonlinear effects. 
The resulting nonlinear tails are robustly analyzed by numerically solving the wave equation in double-null coordinates using finite-difference methods, which efficiently capture the system’s nonlinear dynamics.
As a result, we find that the nonlinear tails closely follow the corresponding linear decay profiles.
The intermediate-time decay rate of the tails is independent to the details of the source or initial data. 
This universality indicates that tail signals alone cannot unambiguously discriminate between linear and nonlinear dynamics of massive fields.
Whereas, quadratic quasinormal modes offer a way to probe the nonlinear effects inherent in massive fields.
Although our study focused on a massive scalar field for simplicity, we believe the insights gained are applicable to a broader range of massive perturbations.
As next-generation gravitational-wave detectors reach high signal-to-noise ratios, extending this framework to massive gravitational perturbations of rotating black holes may enable the direct investigation of the tail signals in black hole ringdowns, offering fresh insights into strong-field dynamics.

\section*{Acknowledgements}
This work is supported in part by the National Key R\&D Program of China, Grant No. 2020YFC2201300, No. 2021YFC2203001, and the National Natural Science Foundation of China, Grants No. 12035016, No. 12375058,  No. 12361141825, No. 12447182, and No. 12575047.

\bibliography{mybibfile}

\end{document}